\documentclass[manuscript]{emulateapj}

\usepackage{overpic}
\usepackage{multirow}
\slugcomment{To appear in the Astrophysical Journal}
\shorttitle{BLAST: FIR Measurement of Star Formation History}
\shortauthors{Pascale et al.}

\newcommand{\degree}{\ensuremath{^\circ}}

\newcommand{\blast}{{BLAST}}
\newcommand{\fidel}{{FIDEL}}

\newcommand{\um}{{$\mu$m}}
\newcommand{\up}[1]{\ensuremath{^{\rm{#1}}}}

\begin{document}

\title{BLAST: A Far-Infrared Measurement of the History of Star Formation}

\author{
Enzo~Pascale$^{1}$,
Peter~A.~R.~Ade$^{1}$,
James~J.~Bock$^{2}$, 
Edward~L.~Chapin$^{3}$,  
Mark~J.~Devlin$^{4}$, 
Simon~Dye$^{1}$,
Steve~A.~Eales$^{1}$,
Matthew~Griffin$^{1}$,
Joshua~O.~Gundersen$^{5}$, 
Mark~Halpern$^{3}$,
Peter~C.~Hargrave$^{1}$, 
David~H.~Hughes$^{6}$, 
Jeff~Klein$^{4}$,
Gaelen~Marsden$^{3}$, 
Philip~Mauskopf$^{1}$, 
Lorenzo~Moncelsi$^{1}$,
Henry~Ngo$^{3}$,
Calvin~B.~Netterfield$^{7,8}$, 
Luca~Olmi$^{9,10}$, 
Guillaume~Patanchon$^{11}$, 
Marie~Rex$^{4}$, 
Douglas~Scott$^{3}$,
Christopher~Semisch$^{4}$, 
Nicholas~Thomas$^{5}$,
Matthew~D.~P.~Truch$^{4}$, 
Carole~Tucker$^{1}$,
Gregory~S.~Tucker$^{12}$, 
Marco~P.~Viero$^{8}$ \&
Donald~V.~Wiebe$^{3,8}$ }

\altaffiltext{1}{ 
School of Physics \& Astronomy, Cardiff University, 5 The Parade,
Cardiff, CF24 3AA, UK; {\url{enzo.pascale@astro.cf.ac.uk}}.  }

\altaffiltext{2}{ 
Jet Propulsion Laboratory, Pasadena, CA 91109-8099, USA.}

\altaffiltext{3}{ 
Department of Physics \& Astronomy, University of British Columbia,
6224 Agricultural Road, Vancouver, BC V6T 1Z1, Canada.}

\altaffiltext{4}{  
Department of Physics \& Astronomy, University of Pennsylvania, 209
South 33rd Street, Philadelphia, PA, 19104, USA.  }

\altaffiltext{5}{ 
Department of Physics, University of Miami, 1320 Campo Sano Drive,
Coral Gables, FL 33146, USA.}

\altaffiltext{6}{ 
Instituto Nacional de Astrof\'isica \'Optica y Electr\'onica (INAOE), 
Aptdo. Postal 51 y 72000 Puebla, Mexico.}

\altaffiltext{7}{ 
Department of Astronomy \& Astrophysics, University of Toronto, 50
St. George Street Toronto, ON M5S~3H4, Canada.}

\altaffiltext{8}{ 
Department of Physics, University of Toronto, 60 St. George Street,
Toronto, ON M5S~1A7, Canada.}

\altaffiltext{9}{ 
University of Puerto Rico, Rio Piedras Campus, Physics Dept., Box
23343, UPR station, Puerto Rico 00931.}

\altaffiltext{10}{ 
INAF, Osservatorio Astrofisico di Arcetri, Largo E. Fermi 5, I-50125, Firenze, Italy.}

\altaffiltext{11}{ 
Universit{\'e} Paris Diderot, Laboratoire APC, 10, rue Alice Domon et
L{\'e}onie Duquet 75205 Paris, France.}

\altaffiltext{12}{ 
Department of Physics, Brown University, 182 Hope Street, Providence,
RI 02912, USA.}

\begin{abstract}
We directly measure redshift evolution in the mean physical properties (far-infrared luminosity, temperature, and mass) of the galaxies that produce the cosmic infrared background (CIB), using  measurements from the Balloon-borne Large Aperture Sub-millimeter Telescope (BLAST), and {\em Spitzer} which constrain the CIB emission peak. 
This sample is known to produce a surface brightness in the \blast\ bands consistent with the full CIB, and photometric redshifts are identified for all of the objects.  We find that most of the 70\,\um\ background is generated at $z \lesssim 1$ and the 500\,\um\ background generated at $z \gtrsim 1$.
A significant growth is observed in the mean luminosity from $\sim10^9$--$10^{12}$ L$_\odot$, and in the mean temperature by 10\,K, from redshifts $0<z<3$.
However, there is only weak positive evolution in the comoving dust mass in these galaxies across the same redshift range.
We also measure the evolution of the far-infrared luminosity density,
and  the star-formation rate history for these objects, finding good agreement with other infrared studies up to $z\sim1$, exceeding the contribution attributed to optically-selected galaxies.

\end{abstract}

\keywords{cosmology: observations --- cosmology: diffuse radiation ---
  submillimeter --- galaxies: evolution --- galaxies: starburst}
 
\section{Introduction}     \label{sec:intro}

The spectrum of the diffuse extragalactic background radiation
reflects the physical processes that have dominated  the
evolution of structure in the Universe.
The Cosmic Microwave Background, the relic
radiation of the early Universe, is the dominant contribution.

The second most important component consists of two broad peaks at
around 200\,\um\ and 1\,\um\ which carry roughly equal energy, and
are presumably associated with light emitted by stars throughout
cosmic time. In the far-infrared (FIR), the isotropic Cosmic
Infrared Background (CIB) was first detected with the FIRAS 
\citep{puget96, fixsen98} and DIRBE \citep{hauser98} instruments on board the COBE satellite. 
It consists of thermal emission from warm dust which enshrouds
star-forming regions in galaxies; thus, a detailed account of the CIB
will by necessity include constraints on the history of star formation
in the Universe \citep[see also][]{dwek98, hauser01}.  In strong contrast to the emission at $\approx 1$\,\um, the warm dust is optically thin, so its intensity reveals the
entire mass without assumptions or corrections.  In this paper we
resolve the contribution to the CIB as a function of redshift.  This
constitutes a measurement of the FIR history of the Universe which is
closely related to the history of star and galaxy formation.

Large fractions of the CIB have been resolved into contributions from individual sources at 24\,\um, 850\,\um, and 1.1\,mm\ \citep[][and references therein]{lagache04}, but the background is several tens of times smaller
at these wavelengths than it is at its peak. 
Near the peak, current and anticipated FIR and submillimeter experiments, including 
the SPIRE instrument \citep{grif03} on the {\sl Herschel} satellite, cannot resolve the CIB into individually detected sources because of confusion arising from the
finite instrumental angular resolution.

It is, however, possible to
study the {\it average} contribution that a given class of objects makes to
the background with a stacking analysis consisting of calculating
the covariance of known source positions with maps from these
experiments. This approach has been successfully used by many authors
\citep[e.g.][]{dole06, dye07, serjeant08}.
Individual sources brighter than $60\,\mu$Jy detected in deep surveys
with the Multiband Imaging Photometer for {\sl Spitzer\/} (MIPS) at
24\,\um\ resolve most of the CIB at 24\,\um\ \citep[${\sim}\,70$\%,][]{papovich04}. Stacking analyses have shown that MIPS sources consitute a large fraction of the background at 70 and 160\,\um\ \citep[$\sim 80$\%, ][]{dole06}. 
The bulk of this emission is from comparatively low redshift galaxies, while at 850\,\um, and 1\,mm the CIB is dominated by massive star-forming galaxies at higher redshifts ($z \approx 2.5$). Besides, \citet{dye07} determine that these extreme galaxies contribute relatively little emission to the important
region near the CIB peak which forms the transition between plentiful nearby
galaxies and rarer distant sources of the submillimetre regime.

\blast\ has observed an 8.7\,deg$^2$ region encompassing the Extended
Chandra Deep Field South (ECDFS) and Great Observatories Origins Deep
Survey South (GOODS-S). Additional time was spent observing a smaller
0.9\,deg$^2$ area at the center of this field, though the coverage is
still broader than most of the other deep data available at other
wavelengths. We refer to these fields as \blast\ GOODS-S Wide
(BGS-Wide) and \blast\ GOODS-S Deep (BGS-Deep) respectively. We note
that the variance of the BGS-Deep maps are dominated by the confusion from point sources arising from the 36\arcsec--60\arcsec\ \blast\ beams, rather than instrumental
noise.

Observing in three broad bands centered at 250, 350, and 500\,\um,
\blast\ provides a new data set of unique size, resolution, and
spectral coverage to study the properties of the CIB on the
submillimeter side of its maximum. Together with MIPS observations at
70\,\um, it is now possible to constrain the CIB without assumptions
about any of the physical properties of the sources contributing to
it.

The possibility of determining number counts, spectral energy
distributions, and clustering properties of submillimetre sources directly
from correlations {\it within} the maps, without the requirement to
first extract a catalog of point sources, was pointed out by \citet{knox01}. The approach has the important advantage of avoiding flux
and number density biases associated with the process of catalog
selection.
\citet{devlin09} and \citet{patanchon09}  use this
approach to estimate the number counts of extragalactic sources in the \blast\ bands across nearly five orders of magnitude in flux density, and \citet{viero09} measure the clustering of BLAST sources via the power spectrum.

\citet{devlin09}, \citet{marsden09}, and this work extend that idea by
measuring the covariance with external data sets, but again without forming a BLAST catalog, which is insted studied by \citet{dye09} who find MIR
and radio counterparts for a sample of bright \blast\ sources and
measure their redshift distribution.
Using this stacking technique, it is found that the total intensity emitted by MIPS 24\,\um-selected sources at the \blast\ wavelengths is compatible with the FIRAS
measurements of the CIB intensity within the FIRAS $1\sigma$
uncertainty, which is $\approx$ 25\%.  The 24\,\um-selected source
redshift distribution has a median of $z \approx 0.9$ (see \S \ref{sec:redshifts}), and it is certainly possible that a population with fainter 24\,\um\ flux
densities is missing, possibly associated with the submillimeter
galaxies detected by SCUBA and MAMBO.

We extend this work by finding the covariance with catalogs which contain
redshift information. We present a
detailed analysis of all sources that contribute significantly to the
CIB at 70\,\um, 250\,\um, 350\,\um, and 500\,\um\ as a function of
redshift, even those well below the confusion limit. From this we
derive the evolutionary history of the CIB. The results from these
five analysis papers will be used by Chapin et al. (in preparation) to
infer the evolution of the FIR luminosity function.

A portion of the BGS-Deep region has been surveyed at many wavelengths
from which we have derived an almost complete set of redshifts ($\sim
95$\%) for MIPS sources.  By exploiting \blast\ and MIPS 70\,\um\ data,
we estimate the mean physical parameters for these galaxies
(FIR luminosities, temperatures, and dust masses), and the
total FIR luminosity evolution and star formation history of the
Universe. Our results are in good agreement with existing
measurements.  We find that stars form in optically obscured dusty
galaxies at rates $\sim 3$ times larger than those previously
estimated from optical data at redshifts up to $z \sim 1$\@. We also estimate
the variation of the total comoving dust mass density with redshift, finding only moderate evolution, within the
experimental uncertainties, over the range $z \lesssim 3$.

A flat cosmological model with $\Omega_\Lambda = 0.7$, $h = 0.7$ has been used throughout.

\begin{figure}[!t]
\centering \includegraphics[width=0.8\linewidth, angle=270]{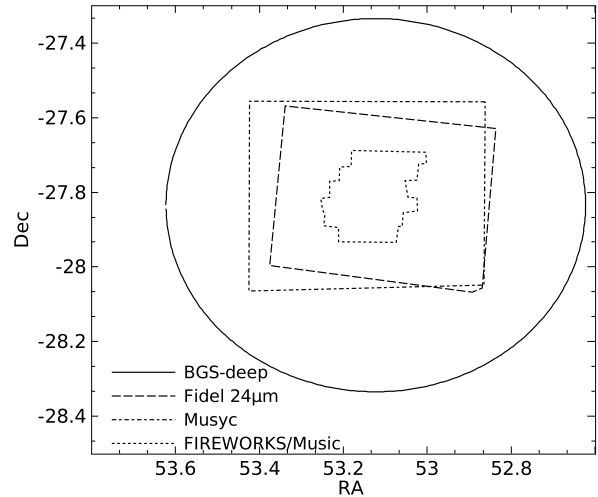}
\caption{Extent of the catalogs in the BGS-Deep region observed with \blast\@.
The area used for the stacking analysis is the intersection of the \fidel\ and MUSYC regions and extends over 590\,arcmin$^2$.
}\label{fig:catalogues}
\end{figure}
\section{Data}     \label{sec:data}
In this section we describe each of the data sets used. Their locations and angular extents  are depicted in Figure~\ref{fig:catalogues}.
\subsection{\blast\ Submillimeter Imaging}
The BGS-Deep region observed with \blast\ covers the central 0.9\,deg\up{2} of a shallower map imaging the full 8.7\,deg\up{2} area in BGS-Wide.

The raw time ordered data (TOD), consisting of a set of voltage time-streams
from each of the \blast\ detectors, are pre-processed \citep{truch, patanchon}.
Any corrupt samples are flagged, and the data are deconvolved using the
instrumental transfer functions.
The TODs are binned into maps using the telescope pointing solution
\citep{pascale} which has an estimated RMS positional error of less than
5\arcsec\ \citep{marsden}. The absolute photometric calibration has an
estimated accuracy of ${\sim}\,10$\%, which is highly correlated across the
\blast\ bands \citep{truch09}. Calibration errors and color corrections are
accounted for in our SED fits.

The maps used in this work have been generated using {\sc optbin} (Pascale
et al.~in preparation).
The algorithm performs common-mode suppression and high-pass filters
the output above the low-frequency ``knee'' of the detector noise. The filtered data streams are then projected onto a
map.  This treatment is appropriate for point source extraction as it
does not distort the data on timescales comparable to the
beam-crossing times.  {\sc optbin} reduces the 100\,hour long BGS-Wide
data-set at 250\,\um\ in one hour on a single--CPU machine, and
it is also used as part of an end-to-end instrumental simulator which
allows Monte Carlo analysis.

Results have been compared and found to be compatible with those
obtained using maps reduced with the more computationally intensive
{\sc sanepic} algorithm \citep{patanchon}, which provides the
least--squares solution to the map-making equation.

\subsection{{\sl Spitzer\/} Data}
Extensive {\sl Spitzer\/} maps cover most of the BGS-Deep and Wide
regions.  The {\sl Spitzer\/} Wide-Area Infrared Extragalactic survey
\citep[SWIRE,][data release~3]{lonsdale03}, provides
$\sim$7.5\,deg$^2$ maps well matched to the BGS-Wide coverage.  We use
the \blast\ data in conjunction with the 70\,\um\ SWIRE data to
constrain the brightness of the peak in the CIB.

Deep imaging of this field is available in the four IRAC (Infrared
Array Camera) bands from the SIMPLE \citep[{\sl Spitzer}'s IRAC and
  MUSYC Public Legacy of the ECDFS,][]{damen09} survey. MIPS data from
the {\sl Spitzer\/} Far-Infrared Deep Extragalactic Legacy (\fidel)
survey are also available. The catalog of 24\,\um\ sources detected in the \fidel\ maps of \citet{magnelli09} is the same as used by \citet{devlin09} and
\citet{marsden09}, and utilizes the IRAC sources detected in SIMPLE as a positional prior.

The faintest 24\,\um\ source in the catalog has a flux density of
13\,$\mu$Jy,  but the detections at $S_{24} < 30\,\mu$Jy should be considered tentative.
The catalog includes 9{,}110 sources
covering 700 arcmin$^2$, with IRAC flux densities (at 3.6, 4.5, 5.8, and 8\,\um)
available for each entry.

\subsection{Optical Photometric Redshifts}

We use five different catalogs of photometric redshifts available in the BGS-Deep region to assign a redshift to the majority of the MIR sources in the \fidel\ catalog.

\citet{brammer08} released two catalogs of photometric redshifts in
the BGS-Deep region using MUSYC \citep[MUltiwavelength Survey by
  Yale-Chile,][]{taylor09} and FIREWORKS \citep{wuyts08} imaging data,
calculated with a new algorithm called {\sc eazy}. {\sc eazy} performs linear
combinations of SED templates to find the best fits to measured
photometry, which in these two cases span the optical ({\sl UBVRI\/})
and the IR ({\sl JHK}), as well as the 4 IRAC bands. They provide
values for the $Q_z$ statistic which indicates the confidence level
associated with the inferred redshifts. As explained therein,
redshifts with $Q_z < 3$ result in a statistical error of $\sigma_z
\equiv \Delta z/(1+z) < 0.1$. Whenever the $Q_z$ statistic is
available, we select redshifts using this cut.  The MUSYC catalog
covers an area of 900 arcmin\up{2} and contains $\sim 17{,}000$ sources,
while FIREWORKS extends over a 140 arcmin\up{2} area and has $\sim
6{,}300$ sources. The typical redshift errors are $\sigma_z \lesssim
0.04$.

The GOODS-MUSIC \citep{grazian06} catalog contains $\sim 15{,}000$
sources in a region overlapping FIREWORKS. About 6.5\% of the
redshifts listed in the catalog are spectroscopic, and the others are
photometric estimates. The photometric redshifts are based on data at
similar wavelengths to those used by \citet{brammer08}, but the
algorithm itself differs from {\sc eazy}. \citet{taylor09} show that the two
sets of redshift estimates are in equally good agreement with the
available spectroscopic redshifts.  However, this agreement does not suggest
 which methodology is more accurate for the
optically dim, high--redshift sources which dominate in the
submillimeter, and for which spectroscopic redshifts are much
rarer. The overall photometric redshift error is estimated to be $\sigma_z
\sim 0.06$ for both catalogs. 

COMBO--17 photometric redshifts \citep{wolf04, wolf08} are available
for the same region of the sky as MUSYC. The 5 broad and 12 narrow photometry
bands from 350 to 930\,nm enable accurate SED measurements,
although the lack of near infrared data limits COMBO--17 capabilities
to estimate redshifts reliably at $z \gtrsim 1.2$. The accuracy of
these redshifts is $\sigma_z \sim 0.01$ for galaxies with $R < 21$,
0.02 for galaxies with $R \sim 22$ and 0.1 for those with $R > 24$,
which is the sensitivity limit of the survey. Therefore, following
other authors \citep[e.g. ][]{lefloch05}, we only use COMBO--17 for
sources with listed $R$ magnitude brighter than 24.

Finally \citet{RR08} provide a catalog of photometric redshifts for
the 24\,\um\ sources detected in SWIRE over a larger 4.56\,deg\up{2}
region within BGS-Wide, but with a shallower depth compared to
\fidel\@.  The optical data ($g'\,r'\,i'$) are combined with the two
shorter IRAC wavelengths, and are used to fit SED templates. The
accuracy in $\sigma_z$ is reported to be 0.035 and is comparable with
the above catalogs.

\section{Redshifts of MIPS Sources}     \label{sec:redshifts}
\begin{figure}[t]
\includegraphics[width=\linewidth]{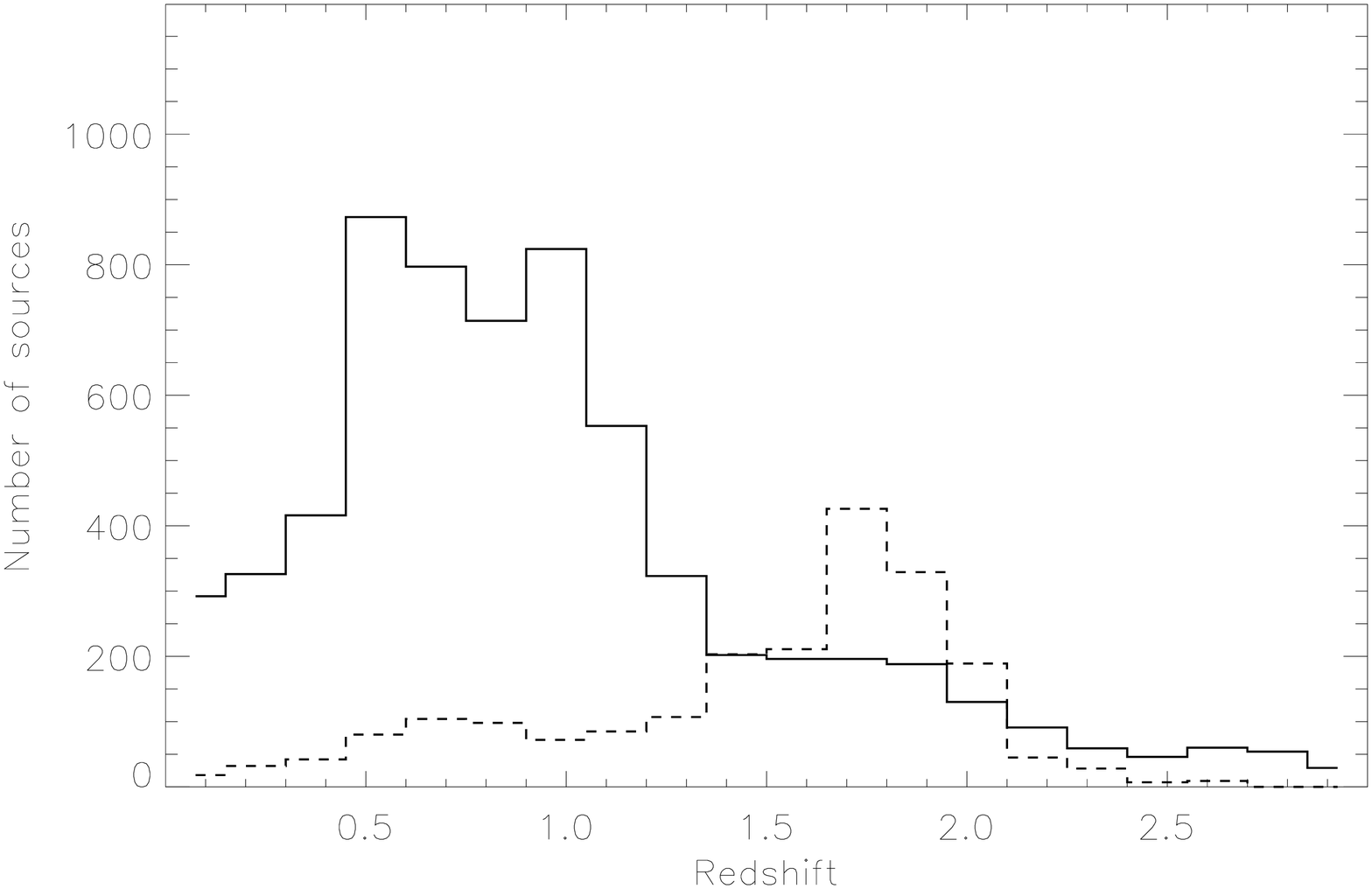}
\caption{The redshift distribution of MIR sources is shown as a solid line for entries having a reliable photometric redshift assigned to them. The distribution peaks between $0.5 < z < 1$ and exhibits a substantial tail up to $z \sim 2$. The dashed histogram shows the distribution of sources with redshifts derived from the IRAC flux density ratios, which tend to cluster at $z \gtrsim 1$.
}
\label{fig:z-dist}
\end{figure}
\begin{figure}[!b]
\includegraphics[width=\linewidth]{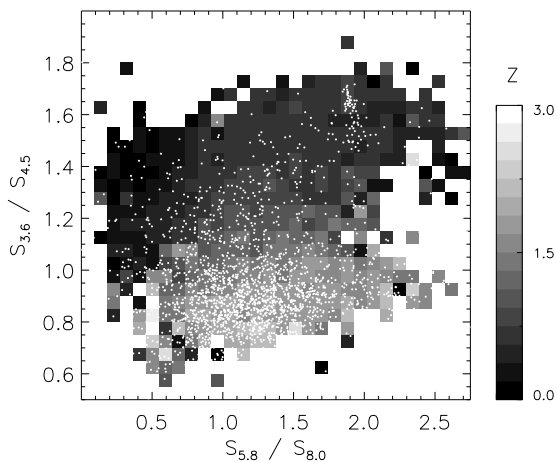}
\caption{ IRAC color-color plot of SIMPLE sources with redshifts.
  Pixels are grey-shaded according to redshift with lighter shades
  corresponding to higher redshifts.  Overplotted as white dots are
  MIR galaxies in the \fidel\ catalog for which no redshift
  information is available.  These points cluster in the bottom region
  of the plane, suggesting that they may be a population of higher
  redshift sources. Redshifts were estimated for these objects
  directly from the means of sources with known redshifts that
  landed in the same bins of this color-color plane}
\label{fig:IRAC_col}
\end{figure}

The 9{,}110 sources in the \fidel\ catalog have positional accuracies
that are significantly smaller than the MIPS PSF (Point Spread
Function) at 24\,\um. This is because the catalog positions are
obtained from the more accurate near infrared (NIR) IRAC catalog.  By
matching the \fidel\ and redshift catalogs we identified redshifts for
$\sim 72$\% of the \fidel\ sources.  We used a search radius of
1\arcsec, but using 1.5\arcsec\ or 2\arcsec\ does not substantially
change the number of identified sources.  Given the surface densities
of sources in the catalogs (MUSYC has one source every
190\,arcsec\up{2} and MUSIC about one every 30\,arcsec\up{2}), we
expect only a few spurious associations within this search radius.

We identified redshifts from the catalogs in the following sequence:
(i) FIREWORKS (18\%); (ii) MUSYC (43\%); (iii) MUSIC (1\%); (iv)
COMBO--17 (10\%), and (v) the catalog of \citet{RR08} (0.5\%).  The
redshift distribution (solid line in Figure~\ref{fig:z-dist}) peaks at
$z \sim 0.8$\@ and shows a high redshift tail which is particularly 
significant in the range $1.5 \lesssim z \lesssim 2$\@.

To determine redshifts for the remaining 28\% of the MIR sources with
no direct counterparts in the catalogs we used a relation between IRAC
colors and redshifts similar to that derived in \citet{devlin09}.
Redshifts from FIREWORKS and MUSYC were matched to IRAC sources in the
SIMPLE catalog. The resulting $\sim$18{,}000 objects were binned in a
look--up table (LUT) as a function of IRAC flux density ratios. The
table is shown in Figure~\ref{fig:IRAC_col} and exhibits a trend of
increasing redshift from top-left through top-right
to bottom-right. Redshifts for
galaxies are assigned based on the mean values of sources that land
in the same bin of the IRAC color-color plane with known redshifts.

\begin{figure}[t]
\includegraphics[width=\linewidth]{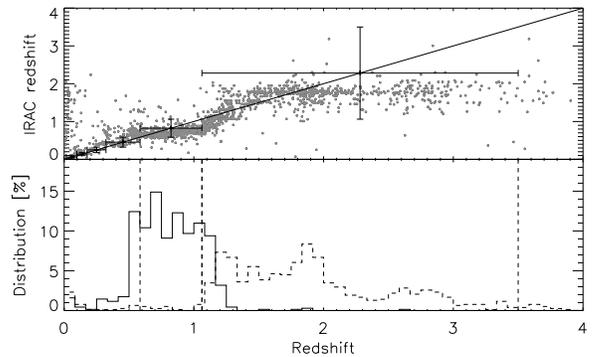}
\caption{Robustness of IRAC redshifts derived from the NIR flux
  density ratios. The top panel compares the IRAC redshifts with the
  catalog of photometric redshifts. Although the relation is not
  linear, its monotonic trend can be used to assign sources to
  redshift bins, provided these are larger than the scatter
  introduced. The redshift binning grid is also shown with the crosses
  indicating the extension of each bin. The distribution of redshifts
  for sources whose redshift lies in the last two bins is shown in the
  bottom panel. The in-bin fraction is 68\% and 90\%\@.}
\label{fig:z-dist-check}
\end{figure}

A comparison between LUT--retrieved and known redshifts for a number
of sources which have not been used to compile the table 
is shown in Figure~\ref{fig:z-dist-check}. 
Although the trend has systematic deviations, as well as a large scatter, the plot exhibits monotonic behavior, and a reliable separation between low and high
redshifts. Two plateaus are evident, one at $0.6 < z < 1.1$ and the other at $z > 1.6$.

We initially chose nine redshift bins spanning $0.016 < z < 3.5$ at
logarithmically spaced intervals. The first three bins are narrower
than the accuracy quoted for the redshift catalogs, so we merged them
to form a single bin 4\% wide in $\Delta z / (1+z)$. We also merged
the last two bins due to the small numbers of sources that landed
within them.  The six final bins are overplotted in the top panel of
Figure~\ref{fig:z-dist-check}.  The redshift distribution for the
sources selected in the last two bins from the IRAC LUT are also
shown. Their widths are 68\% and 90\%, respectively.  IRAC redshifts
are not reliable in the lower redshift bins.

Almost all of the 28\% of the \fidel\ sources with no 
redshifts have flux density ratios populating the
bottom region of the IRAC plane, and their IRAC photometric redshift
distribution (Figure \ref{fig:z-dist}) indicates that they lie
preferentially at $z > 0.5$. This is also confirmed by studying the
distribution of COMBO--17 redshifts for these sources. Although
unreliable in detail, as they are mostly $R \gtrsim 24$, the distribution peaks
in the same redshift range.

In summary, we have obtained or estimated redshifts for 95\% of the
sources from the original MIPS catalog.

\section{Completeness}

The completeness of the flux-limited \fidel\ catalog is investigated
in \citet{devlin09}, finding it to be 63, 80 and 96\% complete at 20,
40, and 80\,$\mu$Jy, respectively.  
Ideally, one would like to
have a similar completeness analysis for each redshift bin used
here. Although it is tempting to apply the same corrections, in
practice one should consider that the catalog is based on IRAC
positions and it would be arbitrary to assume a one-to-one relation
between sources missing in \fidel\ and sources missing in the
underlying SIMPLE catalog. In this work we have therefore ignored any
completeness correction.  However, we do indicate what the expected effect
might be in the analysis and discussions that follow.

\section{Stacking Analysis} \label{sec:stacking}
\begin{deluxetable*}{lcccc}
\tablewidth{0pt}
\tablecolumns{6}
\small
\tablecaption{Stacked Intensities \label{table:stackfluxes}}
\tablehead{ \colhead{$\lambda$} & \colhead{Total} & \colhead{Total\tablenotemark{a} (sources} & 
	\colhead{Total\tablenotemark{b} (sources with}& CIB\tablenotemark{c} \\
	\colhead{}& \colhead{(all sources)}&\colhead{with redshift)}&\colhead{IRAC-z excluded)}&\colhead{}\\
	\colhead{\um} & 
	\multicolumn{4}{c}{----------------------------~~~$\nu\,I_\nu$ [nW\,m\up{-2}\,sr\up{-1}]~~~----------------------------}   }
\startdata
                70         & $5.4\pm 0.3$  & $5.2\pm 0.3$ & $4.6\pm 0.3$ & ... \\
               250         & $8.2\pm 0.5$  & $8.1\pm 0.5$ & $6.1\pm 0.5$ & 
		$10.4 \pm 2.3$  \\
               350         & $4.8\pm 0.3$  & $4.8\pm 0.3$ & $3.3\pm 0.2$ &
		$~5.4\pm 1.6$ \\
               500         & $2.0\pm 0.2$  & $2.0\pm 0.2$ & $1.2\pm 0.1$ & 
		$~2.4\pm 0.6$ \\
\enddata
\tablecomments{Stacked intensities have not been corrected for completeness, and the quoted errors do not include calibration uncertainties. See \citet{marsden09} for the completeness corrected values.}
\tablenotetext{a}{Stacked intensity from every source with a redshift}
\tablenotetext{b}{Stacked intensity from every source with a redshift, excluding IRAC-estimated redshifts (see text in \S \ref{sec:redshifts})}
\tablenotetext{c}{FIRAS measured background \citep{fixsen98}}
\end{deluxetable*}%

Stacking analysis using positional information of sources selected at
different wavelengths is a powerful tool for estimating the
contribution from a given class of objects to the CIB at different
wavelengths. In this paper we have considered the contribution of
MIR-selected sources to the CIB as a function of redshift at \blast\ and
{\sl Spitzer\/}-70\,\um\ wavelengths.  Our analysis follows the method of
\citet{dole06}.  \blast\ maps are whitened to suppress the larger
scales not relevant to a point-source analysis. The filter which makes the noise white is estimated from the smoothed two-dimensional power spectrum of the map. Scales below $\sim 20\arcmin$ are also suppressed. 
The whitening filter
gain is taken into consideration when flux densities are measured.
The region considered conservatively excludes the edges of the
\fidel\ and MUSYC area coverage. This region is described by a
quadrilateral with corners at the following coordinates (J2000):
3\up{h}31\up{m}36\up{s}, $-28\degree 02^\prime 16^{\prime\prime}$;
3\up{h}31\up{m}29\up{s}, $-27\degree38^\prime34^{\prime\prime}$;
3\up{h}33\up{m}16\up{s}, $-27\degree35^\prime00^{\prime\prime}$;
3\up{h}33\up{m}26\up{s}, $-27\degree58^\prime50^{\prime\prime}$. 7{,}280
\fidel\ sources are in this region, compared to the total 9{,}110 listed
in the full catalog.


The means of each map in the stacking region are subtracted, as
required by the stacking formulation.  We then compile a list of
postage-stamp maps centered at the position of each MIR source. These
sub-maps are co-added and the total flux density evaluated using
aperture photometry. Removing the mean locally before stacking is
formally correct; we verified that the standard technique of
subtracting the sky estimated in an annulus around the aperture gives
a compatible result. Whitening the maps removes all significant large
scale structure so that the postage-stamp maps can be co-added without
applying random rotations, as in \citet{dole06}.
Although the SWIRE 70\,\um\ map
was not whitened, the filtering applied by the {\sl Spitzer\/} reduction
pipeline is strong enough to remove large scale fluctuations. We
verified that the retrieved flux densities are the same both by
stacking postage-stamp maps directly, and by alternately rotating them
by 90\degree\ as in the analysis of \citet{dole06}.

The total retrieved flux density can be overestimated if the
distribution of MIR sources is clustered. \citet{marsden09} find this
effect to be negligible for these catalogs and beam-sizes
and it is ignored in the subsequent analysis.

In order to validate the technique and to verify that the analysis
pipeline does not introduce artifacts, stacking was performed on
noiseless Monte Carlo simulations. Sky realizations were generated at
250\,\um\ from model counts \citep{lagache03}. Source flux densities
were drawn from the model and redshifts were assigned from a uniform
distribution spanning $0 < z < 2$. Observed 24\,\um\ flux densities
were assigned to each galaxy using Arp~220 as a template to scale from
the 250\,\micron\ flux densities given the redshifts. Only
24\,\um\ flux densities brighter that 13\,$\mu$Jy were retained in
this list.

We simulated time-stream detector data by scanning the sky
realizations with the actual \blast\ telescope pointing solution. We
then used the \blast\ data reduction pipeline, identical to that for
the real data, to generate maps.

Stacking was performed on the simulated sky realizations and on the
maps obtained from the pipeline, and then compared to the total
250\,\um\ flux density of the sources in the 24\,\um\ mock
catalog. The three measurements were found to agree within 2\%, which
is well within the errors.

A final test was carried out by stacking the \fidel\ sources on the
24\,\um\ {\sl Spitzer\/} map itself. Again, we found that the retrieved flux
density was in good agreement with the total flux density in the list.

The uncertainties are estimated as in \citet{marsden09} and a similar
check for Gaussianity was performed.  
We also verified that stacking against a catalog of randomly selected positions results in a value of zero.

Our stacking region covers an area of only
590\,arcmin\up{2}. Sampling variance can thus play an important role, with
\cite{dole06} suggesting that it can be as large as a factor of 2
(peak-to-peak). Although we do not presently have comparable data in a
different field with which to test this hypothesis, we make a rough
estimate by dividing the present field into 4 separate pieces of
roughly equal areas.  Stacking in each of these regions independently
we find an RMS dispersion of $\sim 15$\% in the retrieved flux
densities, which is fairly correlated among the different \blast\ and
{\sl Spitzer\/} wavebands.  Because of the 4 sub-fields,
the final RMS variation is
expected to be a factor of 2 smaller than this, although we do not
explicitly use this estimate in the error bars later in our analysis.
This is probably only a
lower limit on the sampling variance as this calculation can not
account for scales larger than 1/4 of the map.

Table \ref{table:stackfluxes} lists the results of this analysis. The
70\,\um\ retrieved intensity is compatible with values from
\citet{dole06} and \citet{dye07}. At the \blast\ wavelengths these
values are within 6\% of the ones reported by \citet{devlin09}, and
within 4\% of those in \citet{marsden09}.

The stacked intensities for the 5\% of sources with no redshift
information is negligible compared to the total ($< 3$\%). The table
also lists flux densities retrieved using only sources with redshifts
obtained directly from the catalogs (i.e., leaving out the IRAC redshifts).
These flux densities are 10\%, 30\%, 30\%,
and 40\% lower at wavelengths 70, 250, 350, and 500\,\um,
respectively. This is broadly expected if the excluded sources lie at
high redshift, such that the longer wavelengths are more seriously
affected. It also provides a consistency check for the IRAC redshift
method discussed earlier.

Although the stacked flux density as a function of wavelength is
compatible with the FIRAS detection, given its large error and the
sampling variance, it is still possible that a missing population of
faint 24\,\um\ sources is needed to resolve the CIB completely. This
is mentioned in \citet{marsden09}, and we discuss it further in later
sections. In any case, it is worth noting that existing estimates of the total
resolved CIB are highly uncertain, with FIRAS measurements being affected by
systematics and having a formal error of 25\%.

\begin{figure}[!b]
\includegraphics[width=\linewidth]{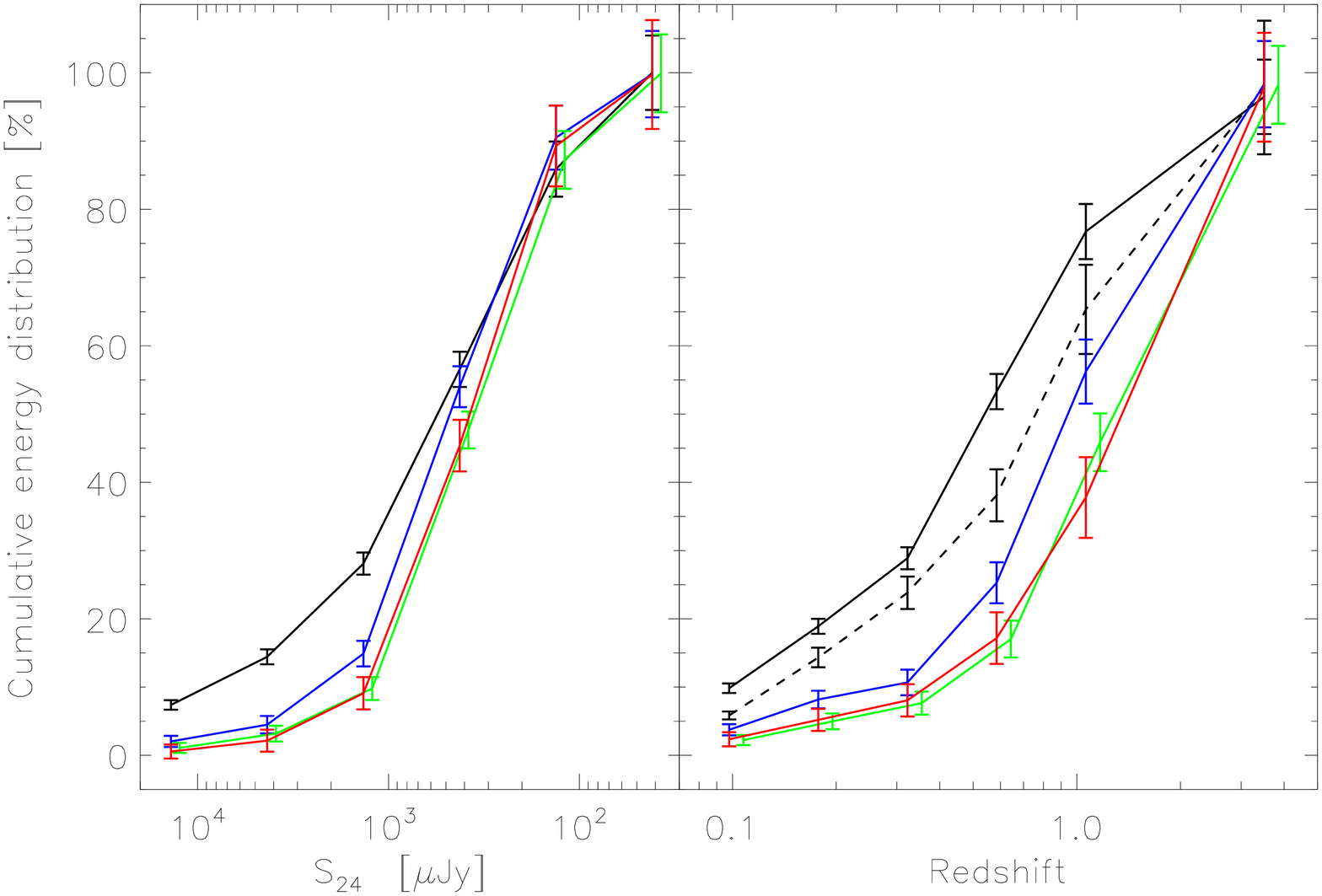}
\caption{The cumulative energy distribution is shown as a function of
  24\,\um\ flux density (left panel) and redshift (right panel) for
  {\sl Spitzer\/} at 70\,\um, in solid black, and \blast\ in blue (250\,\um), green (350\,\um), and red (500\,\um). The 350\,\um\ data are shifted by 10\% to the right in each panel for visual clarity. Each curve shows the integrated
  stacked intensity normalized to the total stacked intensity at each
  wavelength as listed in Table \ref{table:stackfluxes}. Each data point is plotted in correspondence of the rightmost edge of each bin. Different
  wavelengths contribute different amounts to the CIB at different epochs,
  or at different 24\,\um\ flux densities, with most of the
  70\,\um\ background being generated at $z \lesssim 1$ and the
  500\,\um\ background at higher redshifts.
 The cumulative plot as a function of flux density shows a similar
 trend at all wavelengths, suggesting that brighter 24\,\um\ sources
 are likely to be at lower redshifts. The cumulative energy distribution for {\sl Spitzer} at  24\,\um\ is also shown for reference (dashed line in the right panel).}
\label{fig:fluxzdist}
\end{figure}


\section{CIB Redshift Distribution}
The observed CIB is the total dust reprocessed starlight produced in
galaxies at all redshifts and luminosities.  Splitting the
\fidel\ catalog into redshifts bins is an effective way
to study the contribution that each sub-catalog has to the total
intensity. The redshift bins used (see \S \ref{sec:redshifts}) were
chosen to be at least $\sigma_z$ or larger in $\Delta z/(1+z)$, and
the width of the last two bins ensures that the errors in the IRAC
redshifts do not result in sources being placed in the wrong bin
(Figure~\ref{fig:z-dist-check}). The mean stacked intensity in each
bin is listed in Table~\ref{table:stackresults}.

The cumulative distribution of stacked intensities is shown in
Figure~\ref{fig:fluxzdist} as a function of both 24\,\um\ flux density
and redshift, and it is normalized to the total intensity listed in
Table~\ref{table:stackfluxes}. We prefer to show
this normalization rather than normalizing to the
FIRAS measurements because: i) there is no 70\,\um\ FIRAS detection and
the most accurate detections of the CIB at this wavelength are lower
limits; ii) the FIRAS detections in the \blast\ bands have large
uncertainties; and iii) the \blast\ survey is affected by sampling
variance, making it difficult to compare with an all sky survey.
\begin{figure*}[t]
\centering
\includegraphics[width=\linewidth]{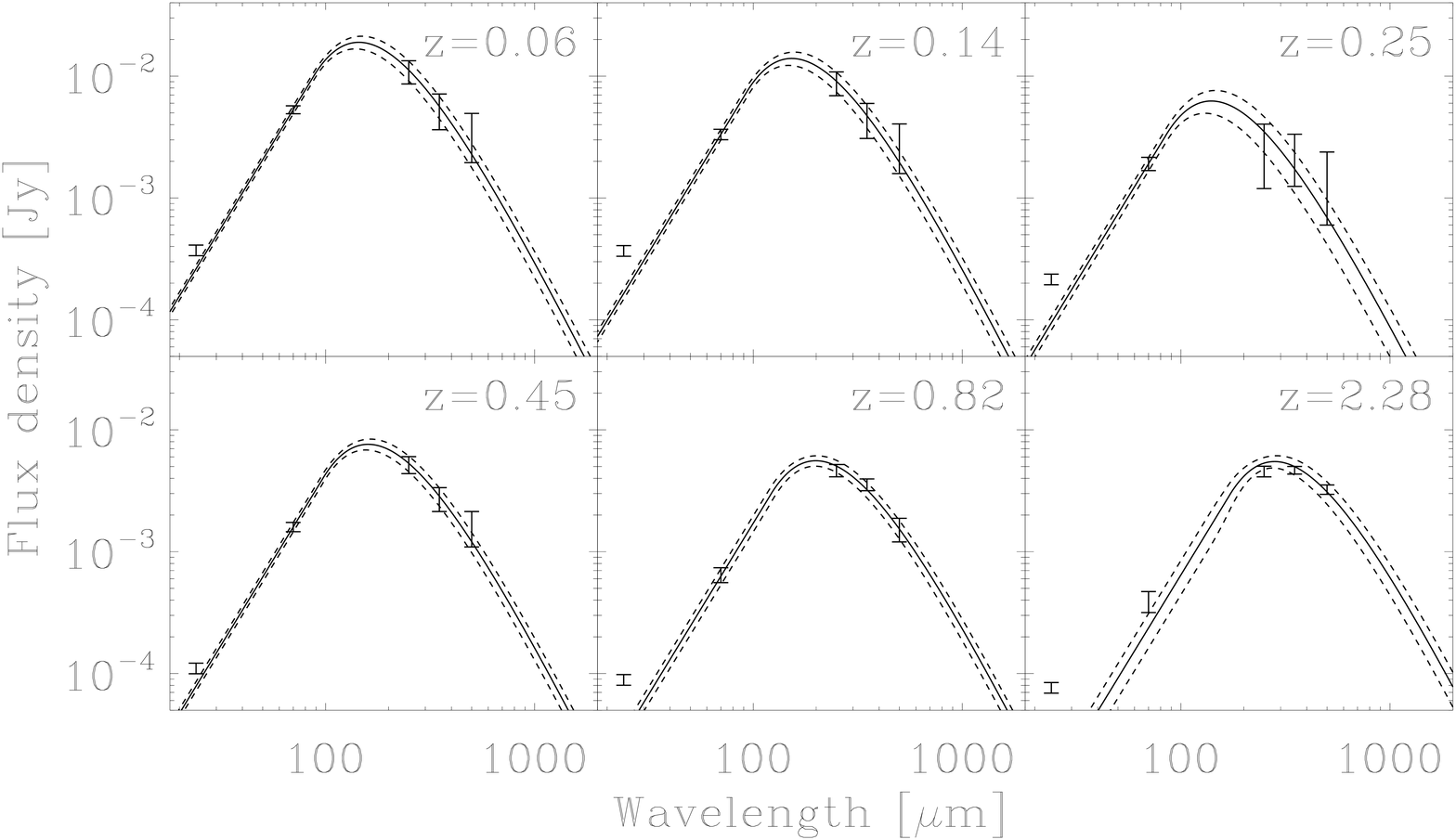}
\caption{SED fitting of the average flux densities measured in different
redshift bins.  The points with error bars are from \blast\ (color-corrected 250, 350, and 500\,\um), and {\sl Spitzer} 70\um\@.  
The solid lines are the best fits at each redshift, with the 68\% confidence levels indicated by the dashed lines. The
fit accounts for the finite \blast\ bandwidths and for the correlated
calibration uncertainties.  The model template is a modified greybody
with an emissivity law $\beta = 1.5$ and a power law $\nu^{-\alpha}$
replacing the Wien part of the spectrum. Two templates were used: one
with $\alpha = 2.8$, describing Arp~220's spectrum, and one with
$\alpha = 1.8$ which is better matched to M82. Both models are
compatible with the data, although for clarity only the $\alpha = 2.8$ fit is
shown here. 
The 24\,\um\ average flux densities of the sources in the catalog are also shown in each panel for completeness, but are not used in the fits because these fluxes come from regions which are not in thermal equilibrium.}
\label{fig:sedfit}
\end{figure*}

A similar qualitative behavior is seen in both panels of
Figure~\ref{fig:fluxzdist}, with the stacking measurements at shorter
wavelengths dominating the brighter 24\,\um\ flux densities and the
lower redshifts. By redshift 1.1, about 75\% of the CIB is generated
at 70\,\um, 55\% at 250\,\um, 45\% at 350\,\um, and 40\% at
500\,\um\@. Similar results are found by \citet{marsden09} who
estimate these fractions to be 50, 45, and 40\% at 250, 350, and
500\,\um, respectively (although $z = 1.2$ is the dividing redshift in their study). The
relatively small discrepancy at 250\,\um\ can be accounted for by the different way
redshifts are assigned. In this work, most redshifts have been
retrieved from catalogs, while \citet{marsden09} only uses an approximate
statistical division into two redshift bins using IRAC colors.

\citet{wang06} stacked  $H$-band and IRAC selected sources
 in the 850\,\um\ map of the GOODS-N field. They find that about 70\% of the total retrieved flux originates at redshifts $z < 1.5$.  This is about 30\% of the FIRAS measurement, suggesting that a large fraction of the CIB at this wavelength may be generated at higher redshifts. \citet{serjeant08} reach similar conclusions stacking MIR and NIR selected sources.

It is worth mentioning that the fractions of the resolved background
depend on the normalization adopted. If FIRAS is used instead, all of
these fractions change by as much as 20\% at the \blast\ wavelengths
(not accounting for the quoted measurement uncertainties).

Compared to 70\,\um, the 500\,\um\ cumulative distribution shows no
evidence of convergence in the highest redshift bin, suggesting that a
population of faint 24\,\um\ sources is missing in order to fully resolve the
CIB at these longer wavelengths\@.

\section{Average Physical Parameters}     \label{sec:fitting}
The average flux densities at 70, 250, 350, and 500\,\um\ from
stacking (total stack divided by the number of contributing sources)
were used to fit SEDs in each redshift interval to estimate the
average FIR luminosities (from 8 to 1000\,\um\ in the rest
frame) of the MIR sources as a function of redshift.

Since the emission mechanism at these wavelengths is thermal, the
natural choice of SED is a modified greybody with some emissivity law:
$ A\,\nu^\beta\, B(\nu, T)$, where $B(\nu, T)$ is the blackbody
spectrum and $A$ its amplitude. Such a model assumes a single dust
temperature, but the reality is far more complex. Temperature
distributions are observed both within single galaxies and among
different sources. As a consequence, the resulting spectrum is better
described by a power law  than a Wien exponential decay at
wavelengths shorter than the peak in the SED.  Indeed, for our data a
modified greybody alone gives a poorer fit to the combined
{\sl Spitzer\/} 70\,\um\ and
\blast\ data points as redshift increases; this is because the
70\,\um\ channel samples shorter rest-frame wavelengths.

Instead of a blackbody curve modified purely on the Rayleigh Jeans side,
we instead used a modified greybody with a
fixed emissivity index $\beta = 1.5$, together with a power law decay
$\nu^{-\alpha}$ at short wavelengths to prevent the high-frequency SED
from falling exponentially.  The exponent $\alpha$ is chosen by
fitting the SEDs of two often used galaxies: the Ultra
Luminous IR Galaxy (ULIRG, $L> 10^{12}\,{\rm L}_\odot$, $\alpha = 2.8$)
Arp~220, and the Luminous IR galaxy (LIRG, $L > 10^{11}\,{\rm L}_\odot$,
$\alpha = 1.8$) M82\@. The former is a merging system, and the latter
is a starburst probably triggered by tidal encounters with the nearby spiral
M81\@.  Our fitting procedures estimate the amplitude of the template and the
temperature, keeping $\alpha$ and $\beta$ fixed. Both SEDs fit the
four data points at each redshift, as shown in Figure~\ref{fig:sedfit}
for $\alpha = 2.8$; the best-fit quantities are listed in Table~\ref{table:stackresults}\@.

\begin{figure}[!t]
\includegraphics[width=0.9\linewidth]{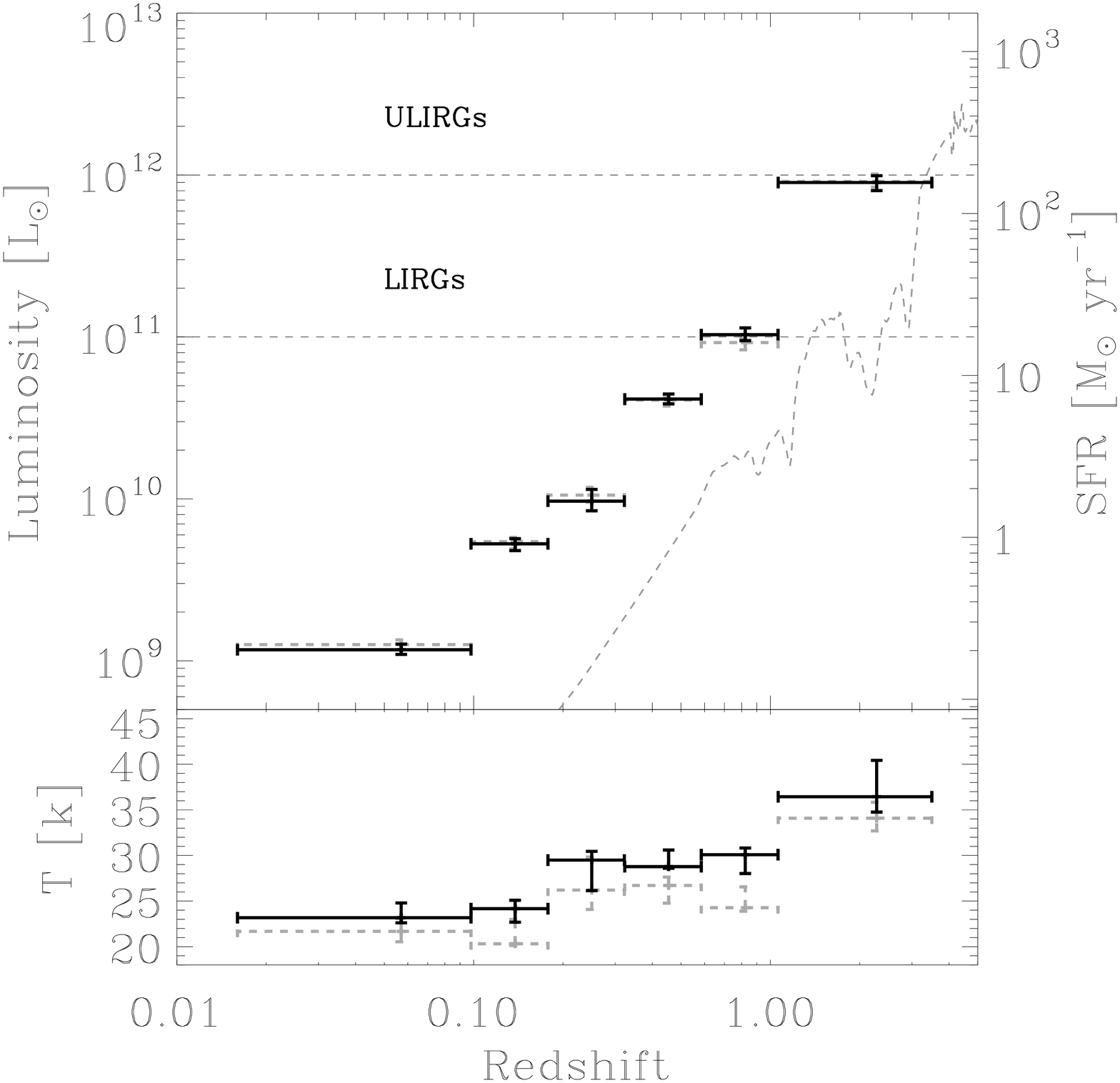}
\caption{The mean rest-frame luminosities, and temperatures  
  of the MIR \fidel\ sources are plotted against redshift.  In
  the top panel, the mean FIR luminosities are estimated from
  the $\alpha = 2.8$ (black points) and $\alpha = 1.8$ (grey,
  dashed points) SED fits. There is little difference between the two,
  confirming that the estimated luminosities are a weak function of
  the template used. The grey dashed line provides an estimate of selection
  arising from a 24\,\um\ flux density limit of 20\,$\mu$Jy (63\%
  completeness limit of the catalog). The bottom panel shows rest-frame
  temperatures for both templates. A difference of $\sim 5$\,K is
  observed between the two, with temperatures apparently increasing with
  redshift.}
\label{fig:LT}
\end{figure}

The estimate of FIR luminosity (the integral from 8 to
1000\,\um\ of the SED in the rest-frame) is a weak function of
the template (Figure~\ref{fig:LT}), emphasizing that the model need
only provide a reasonable interpolation of the data around the peak of
the rest-frame FIR emission. The retrieved luminosities increase
steeply with redshift.  The top panel in Figure~\ref{fig:LT} plots the
mean FIR luminosity as a function of the redshift.  It is in
good agreement with a similar result found by \citet{lefloch05}, who
studied the evolution of FIR luminosities for the same
population of 24\,\um\ sources, although with a higher MIR flux
density limit of 80\,$\mu$Jy\@.  The figure provides an estimate of
the selection effect arising from the flux density limit of the
\fidel\ catalog, which we assume is 20\,$\mu$Jy, the 63\% completeness
limit.  The calculation uses template models for ``normal'' and
``starburst'' galaxies from \citet{lagache03}, co-added after first
normalizing to the same FIR luminosity.

As shown in \citet{kennicutt98}, the FIR emission is tightly
correlated with star formation activity (Star Formation Rate, or SFR),
under the assumption that star formation is completely optically
obscured by dust and that the fraction of infrared luminosity
dominated by AGN heating is negligible\footnote{\citet{devlin09} show
  that the AGN contribution to the CIB at \blast\ wavelengths is about
  7\%, and it is small enough to be ignored.}.  \citet{kennicutt98} shows that the relation between FIR luminosity and dust obscured SFR is then:
\[
 {\rm SFR} [{\rm M}_\odot\, {\rm yr}^{-1}]
 = 1.728 \times 10^{-10} \,L \,\,[{\rm L}_\odot].
\]
We use this relationship to calibrate the SFR axes of the luminosity plots. 

From Figure~\ref{fig:LT}, we observe that at low redshifts ($z
\lesssim 0.6$) the MIR population is dominated by moderate IR emitters
($L \lesssim 6\times 10^{10}\,{\rm L}_\odot$). As the redshift increases,
the volume probed also increases and LIRGs become more common, which
is consistent with many spheroids being formed at about this time ($z
\sim 1$).  This is expected, as SCUBA surveys at 850\,\um\ have
revealed the existence of a large number of ULIRG-type objects,
residing at typically higher redshifts with a median of $z \sim 2.5$
\citep{chapman03}.
\begin{figure}[!t]
\includegraphics[width=\linewidth]{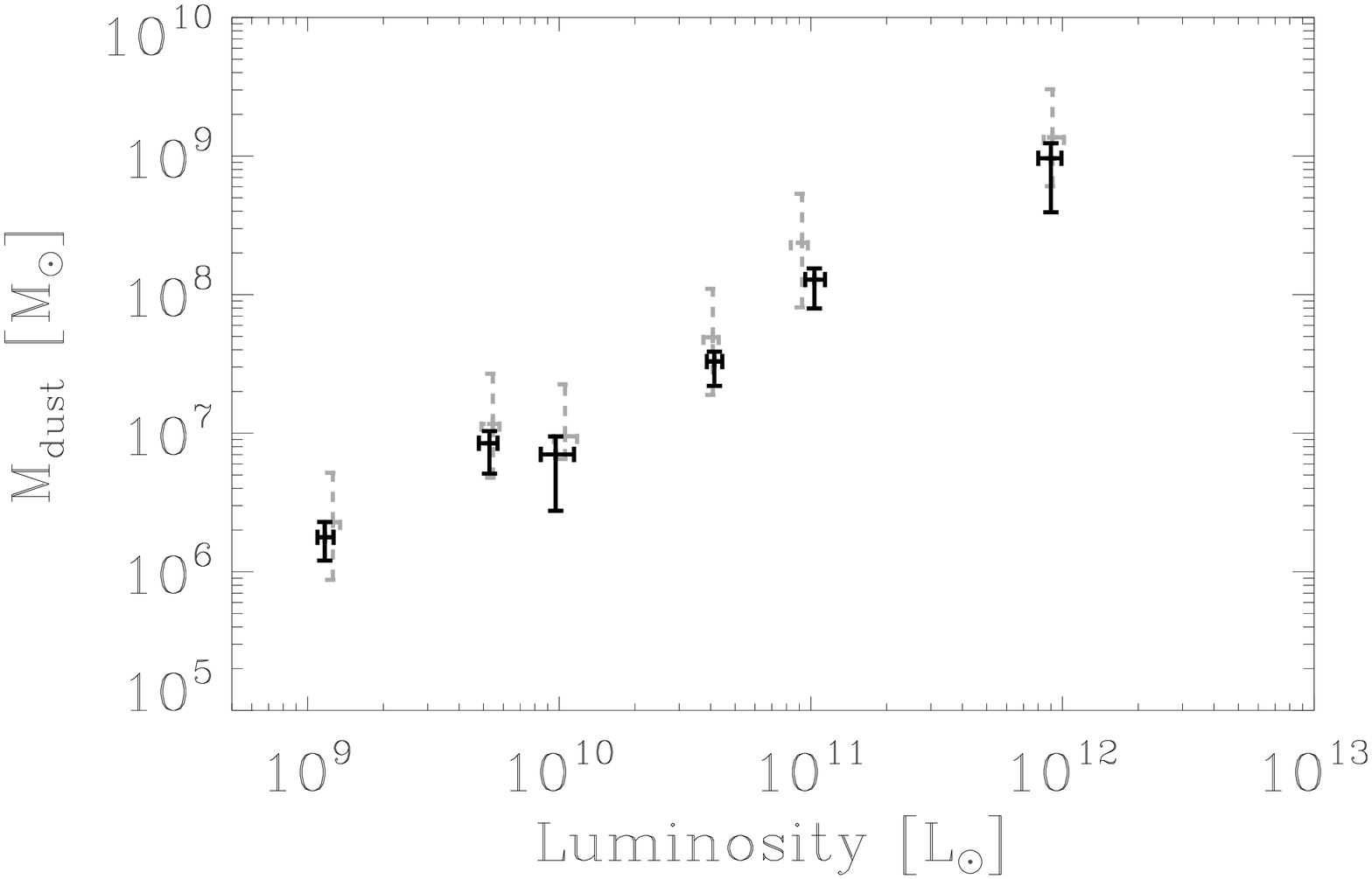}
\caption{The mean dust masses of the MIR \fidel\ sources are plotted against the mean FIR luminosities of Figure~\ref{fig:LT}. These masses are calculated using the temperatures estimated from the $\alpha = 2.8$ (black points), and $\alpha = 1.8$ (grey, dashed points) SED fits. Redshifts are not indicated explicitly, but increase monotonically from left to right.}
\label{fig:ML}
\end{figure}

Stacking analyses cannot be used to study individual sources, only their average properties. Therefore we emphasize once more that the quantities estimated are the average values of the population of MIR sources resolving most, if not all, of
the CIB detected by FIRAS. 
 For instance, the last bin in Figure~\ref{fig:LT} probes a redshift range similar to
the SCUBA galaxies.  However, the mean luminosity is slightly lower
than that typically quoted for SCUBA galaxies, suggesting either that
some of the most luminous objects (and potentially most optically
obscured) are missing from the bin, or that SCUBA galaxies have
luminosities that are above average at those redshifts (which is surely
true for the extreme objects that have been studied so far at 850\,\um).

Our template fitting returns rest-frame temperatures, but these are
more model--dependent than the luminosities. As shown in
Figure~\ref{fig:LT}, the difference in the temperatures determined for
the two templates is about 5\,K across the whole redshift range.

We observe an increase in mean temperature with increasing redshift.  This trend can be
explained by the correlation observed between temperature and
luminosity as discussed by \citet{dunne00}. Detailed modeling beyond
the scope of this paper would be required to determine whether
selection or cosmic evolution also play a role.

The average mass in dust can be estimated from the observed flux
densities \citep{hildebrand83}, $S_{\nu}(\nu_0)$:
\[
 M_{\rm d} = \frac{D_{\rm L}^2(z)}{1+z}\,\frac{S_\nu(\nu_0)}{\kappa\left(\nu_{\rm em}\right) B_\nu\left(\nu_{\rm em}, T\right)}
\]
where $D_{\rm L}$ is the luminosity distance.
\begin{figure}[!t]
\includegraphics[width=\linewidth]{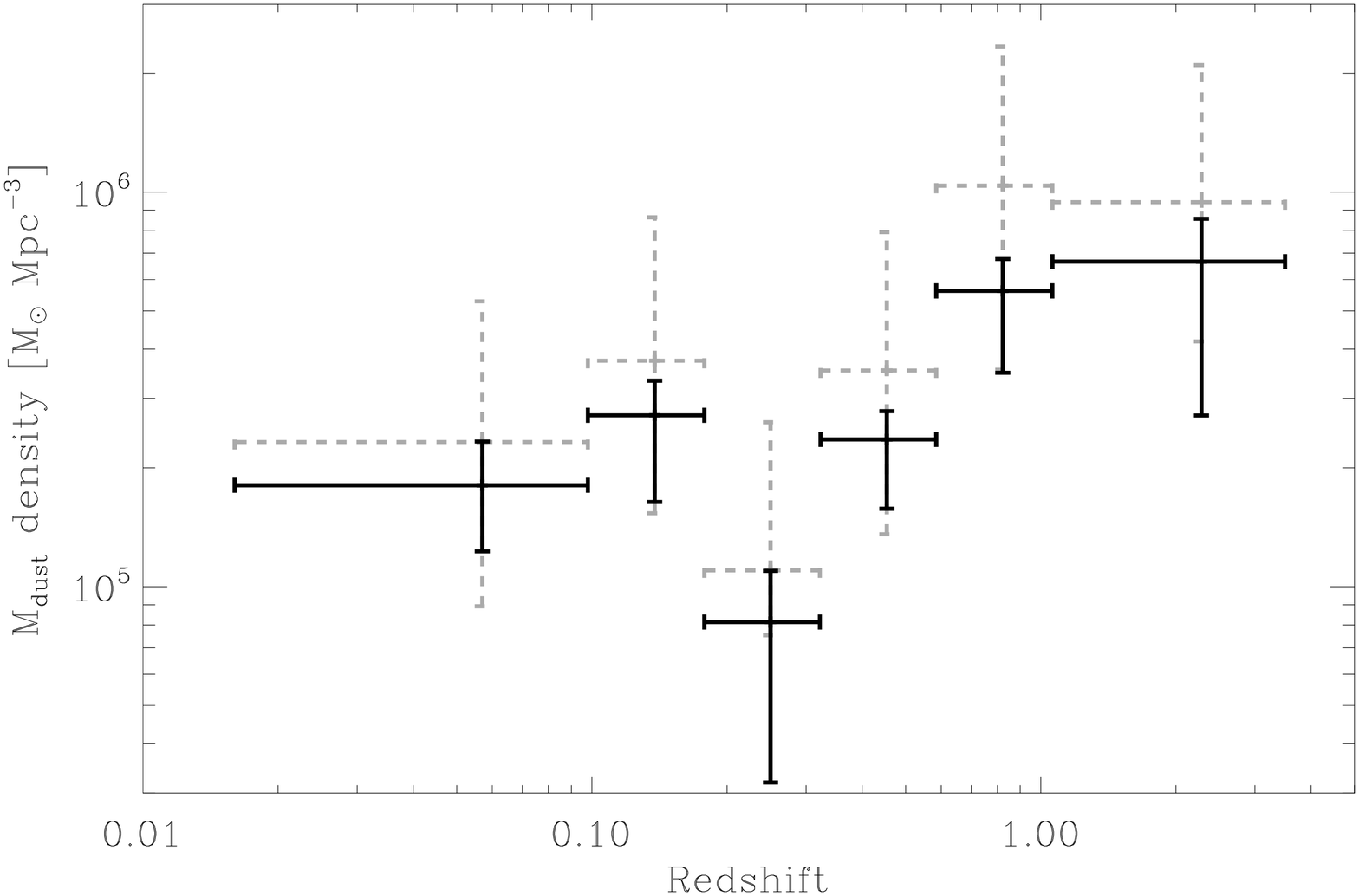}
\caption{Comoving dust mass density vs. redshift. There is only moderate evidence of evolution with redshift.}
\label{fig:mass}
\end{figure}
The mass-absorption coefficient, $\kappa$, is evaluated at 1\,mm
\citep[$\kappa_{\rm 1mm} = 0.1$\,m\up{2}\,kg\up{-1}, ][] {hughes96}, and has
a $\nu^\beta$ dependence.  Any dust mass estimate has to be treated
with caution because: i) $\kappa_{\rm 1\,mm}$ is uncertain within an order
of magnitude \citep{blain02}; and ii) the temperature is largely
uncertain, and model dependent.  We use the flux density at a rest-frame
wavelength of 500\,\um, and temperatures obtained from the two model SED fits. The average dust masses are plotted against the average luminosity in Figure~\ref{fig:ML}.
Given the monotonic relation between FIR luminosities and redshifts, the leftmost and rightmost points in the figure are the lowest and highest redshift bins, respectively.
The dust mass shows a steady increase over two orders of magnitude across the
probed range of redshifts, and FIR luminosities. 

We  observe that the estimates of temperature and dust mass from the fit with the $\alpha = 1.8$ template (M82) have larger errors compared to those obtained from the $\alpha = 2.8$ template (Arp~220). This is due to the flatter spectrum close to the emission peak of the $\alpha = 1.8$ template.

\section{Luminosity and Star Formation Histories}\label{sec:sfh}
\begin{figure}[!t]
\includegraphics[width=\linewidth]{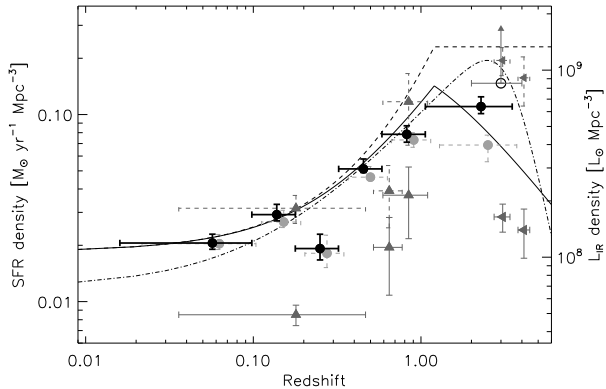}
\caption{The evolution of the total FIR luminosity density up to
  $z=3.5$ from MIR--selected sources is shown as black filled circles. The
  other points are optical-UV detections from \citet{lilly96}
  (up--pointing solid triangles) and \citet{steidel99} (left--pointing
  solid triangles). 
These measurements need to be corrected upward for the effects of dust extinction before they can be compared to detections in the FIR, and
  the corresponding symbols with dashed error bars
  are the extinction--corrected values. The open circle is the lower
  limit of \citet{hughes98} derived for SCUBA galaxies. SFR is shown
  on the left axis and shows good agreement with the optical
  extinction corrected estimates, confirming that most of the
  star-forming activity in the Universe is obscured by dust.  The
  light grey filled circles, with dashed error bars, are calculated
  from the same MIR population, from which the sources with redshifts
  estimated from IRAC colors have been removed, and are shifted by
  10\% toward increasing $z$ for visual clarity.  A luminosity
  function model (not a fit) is plotted as a solid line, and it is
  also evaluated at a 24\,\um\ flux density limit of 20\,$\mu$Jy to
  indicate the effect of the \fidel\ catalog flux density limit
  (dashed line). The data also show good agreement with the model of
  \citet{hopkins06} (dash--dot line). The last redshift bin falls low 
  compared with the SCUBA lower limit (and the extinction--corrected point at 
  similar redshift), and suggests that a fraction of SCUBA galaxies might be a
  population of faint 24\,\um\ sources missing from the
  \fidel\ sample.}
\label{fig:madau}
\end{figure}
Mean rest-frame FIR luminosities and dust masses can be
converted into comoving-volume densities. The conversion
factor applied to each redshift bin is the ratio of the number of
sources in the bin to the comoving-volume probed by the bin, which,
for a Euclidean geometry, we calculate as
\[
 V_{\rm com}(\bar{z}) = \frac{\Omega_{\rm stack}}{3} \left[
	\frac{D_L^3(z_{\rm hi})}{(1+z_{\rm hi})^3} -
	\frac{D_L^3(z_{\rm lo})}{(1+z_{\rm lo})^3} 
\right].
\]
Here, $\Omega_{\rm stack}$ is the solid angle of the stacking region, 
$z_{\rm lo}$ and $z_{\rm hi}$ are the edges of each redshift interval, and
$\bar{z}$ is the bin center.

Interestingly, the comoving dust mass densities plotted in Figure~\ref{fig:mass} show little variation over the redshifts probed. This suggests that the amount of comoving dust associated with the \fidel\ sources is fairly constant as galaxies are built up. \citet{dunne03} found that there was much more dust at $z \gtrsim 1$, although their survey is only sensitive to galaxies with very large dust masses, whereas our stacking analysis is sensitive to galaxies with a wider range of dust masses.

The comoving FIR luminosity densities are shown in
Figure~\ref{fig:madau} and are listed in
Table~\ref{table:stackresults}.  The left vertical axis is converted
into star formation rate density.  We used the template with $\alpha
= 2.8$ to calculate the luminosities.  Also shown is the effect of
excluding sources for which redshifts were obtained from the IRAC
color-color plane.  This potential incompleteness mostly affects the higher redshift bin.
We note once again that the error bars include only the statistical
uncertainties arising from the stacking and calibration and do not
include the effect of sampling variance or incompleteness.
This plot is a complete
representation of the energetics in the CIB generated by MIR sources
at the \fidel\ flux density limit, as 95\% of these galaxies have been
included and the remaining 5\% do not contribute substantially to the
background.
\begin{turnpage}
\begin{deluxetable*}{cccccccclcc}
 \tablewidth{0pt}
\tablecolumns{11}
\tablecaption{Stacking Analysis Results \label{table:stackresults}}
\tablehead{
\colhead{$z_{\rm lo}$} & \colhead{$z_{\rm hi}$} & \colhead{$\nu I_\nu(70\,\mu{\rm m})$} & \colhead{$\nu I_\nu(250\,\mu{\rm m})$} & \colhead{$\nu I_\nu(350\,\mu{\rm m})$} & \colhead{$\nu I_\nu(500\,\mu{\rm m})$} &\colhead{$L_{\rm IR}$\tablenotemark{a}}&\colhead{Temperature\tablenotemark{a}} &\colhead{Dust mass\tablenotemark{a}}&\colhead{$L_{\rm IR}$ density} &\colhead{Sources in bin}\\
\multicolumn{2}{c}{}  & 
\multicolumn{4}{c}{--------------------------- nW\,m\up{-2}\,sr\up{-1} ---------------------------}  &
  10\up{9}${\rm L}_\odot$ & K & 10\up{7}M$_\odot$&10\up{8}${\rm L}_\odot$\,Mpc\up{-3}&} 
\startdata

0.016 & 0.098 & $0.53\pm 0.04$ & $0.31\pm 0.07$ &0.11 $\pm 0.04$ & $0.05\pm 0.02$ &
$1.2_{-0.1}^{+0.1}$&$23.2_{-0.6}^{+1.6}\; [21.7_{-1.1}^{+1.4}]$ &$1.8_{-1.2}^{+2.2}\;[2.3_{1.4-}^{+2.9}]$& $1.19_{-0.09}^{+0.14}$ &115\\
0.098 & 0.177 & $0.49\pm 0.04$ & $0.36\pm 0.08$ &0.13 $\pm 0.04$ & $0.06\pm 0.03$ &
$5.3_{-0.5}^{+0.4}$&$24.2_{-1.5}^{+0.9}\; [20.3_{-0.2}^{+2.7}]$ &$8.4_{-5.1}^{+10}\;[11.6_{-6.9}^{+15.2}]$& $1.69_{-0.13}^{+0.23}$ &169\\
0.177 & 0.322 & $0.54\pm 0.06$ & $0.20\pm 0.11$ &0.13 $\pm 0.06$ & $0.06\pm 0.03$ &
$9.7_{-1.2}^{+1.8}$&$29.5_{-3.3}^{+1.0}\; [26.2_{-2.1}^{+3.6}]$ &$7.0_{-2.8~}^{+9.4~}\;[9.4_{-3.0}^{+13.0}]$& $1.11_{-0.15}^{+0.21}$  &323\\
0.322 & 0.585 & $1.31\pm 0.11$ & $1.19\pm 0.19$ &0.45 $\pm 0.10$ & $0.19\pm 0.06$ &
$41.3_{-2.7}^{+3.1}$&$29.8_{-0.2}^{+1.8}\; [26.7_{-1.9}^{+0.9}]$ &$32.9_{-21.9}^{+38.8}\;[49.2_{-30}^{+61}]$& $2.96_{-0.13}^{+0.39}$ &949\\
0.585 & 1.062 & $1.26\pm 0.16$ & $2.53\pm 0.30$ &1.38 $\pm 0.16$ & $0.42\pm 0.09$ &
$103_{-8}^{+10}$&$30.0_{-2.1}^{+0.7}\; [24.3_{-0.4}^{+2.3}]$ &$128_{-80}^{+154}\;[236_{-156}^{+297}]$& $4.54_{-0.42}^{+0.50}$ &2247\\
1.062 & 3.500 & $1.06\pm 0.19$ & $3.45\pm 0.35$ &2.51 $\pm 0.18$ & $1.22\pm 0.11$ &
$899_{-98}^{+91}$&$36.4_{-1.7}^{+4.0}\; [34.1_{-1.4}^{+1.7}]$ &$936_{-393}^{+1238}\;[1363_{-758}^{+1670}]$& $6.39_{-0.53}^{+0.81}$ &3115 
\enddata
\tablecomments{FIR luminosities, Temperatures and dust masses are the mean values per source in each redshift interval. The FIR luminosity densities are calculated from the mean FIR luminosities as described in Section~\ref{sec:sfh}.}
\end{deluxetable*}
\end{turnpage}
The luminosity density plot shows the luminosity integrated over the
comoving volume. A na\"{i}ve model is overplotted (not fitted) and is
derived from the \citet{saunders90} luminosity function (LF) form
for {\sl IRAS} galaxies,
\[
 	\Phi_{\rm IR}(L) = \phi^*_{\rm IR}
	\left(\frac{L}{L^*_{\rm IR}}\right)^{1-\alpha_{\rm IR}}
	\exp{\left[
	-\frac{1}{2\sigma^2_{\rm IR}}\log^2\left(1+\frac{L}{L^*_{\rm IR}}\right)\right] 
	}.
\]
The coefficients adopted for this model are from \citet{lefloch05} for
a fit to the local FIR LF.  The quantity plotted is then
\[
 \frac{dL}{dV_{\rm com}} = \int^\infty_{L_{\rm min}} L\; \Phi_{\rm IR}\left(\frac{L}{f(z)}\right)\, d\log{L}.
\]
$L_{\rm min}$ is set to zero in one case (dashed curve) to show the
total FIR luminosity density predicted by this model. The solid curve
was calculated for $L_{\rm min}$ estimated assuming a 24\,\um\ flux
density of 20\,$\mu$Jy (63\% completeness) and our fitting template,
and shows the effect of the flux density limit in the \fidel\ catalog
(higher flux density limits correspond to larger values of $L_{\rm
  min}$).
The evolution was chosen to be $f(z) = (1+z)^{3.2}$ up to $z = 1.2$,
and constant at higher redshifts.
This na\"{i}ve modeling shows good agreement with the data and
qualitatively demonstrates the effect of the flux density limit of the
catalog.  As expected, the only substantial change is to the highest
redshift point.

The  FIR luminosity densities estimated here show good
agreement with similar results obtained from optical
detections \citep{lilly96} at $z \lesssim 1$, after a correction for dust extinction has been applied by the authors. Optical measurements are sensitive to the fraction of photons not absorbed by dust. Therefore dust extinction needs to be taken into consideration in order to estimate the total star fromation activity, and to compare with measurements obtained in the FIR\@. 
In Figure \ref{fig:madau} we also plot the empirical model of \citet{hopkins06} obtained from a parametric fit to SFR densities measured at wavelengths spanning from the radio to the UV, which shows good agreement with our measurements. 

As the MIR sources resolve most, if not all, of the CIB, at its emission peak,
this is the first direct measurement of its energetics and of the
dust obscured star formation history of the Universe. Although the
agreement with previous measurements is remarkably good, previous
studies have had to rely on modeling of dust obscuration and SEDs to
evaluate FIR luminosities. With the combination of \blast\ and
{\sl Spitzer\/}, we have been able to directly probe this quantity
with very few assumptions about the underlying physical details.

Comparing our measurements with the optical measurements (the triangle
symbols with solid error bars in Figure \ref{fig:madau}), we find that
the fraction of light from young stars hidden by dust is about $70$\%
in the range $0.1 < z < 1$. 

The highest redshift bin, which coincides with the redshifts at which
SCUBA galaxies are typically found, is the most affected by the
\fidel\ flux density limit. Comparing with the lower limit on the SFR
density for SCUBA galaxies of \citet{hughes98}, it appears as if some
fraction of this population is missing from the \fidel\ catalog, which
does not contain large fractions of ULIRG-type objects, as discussed in
the previous section. It is possible that a significant fraction of
SCUBA galaxies have fainter 24\,\um\ flux densities than the threshold
we used for the \fidel\ catalog, although we note that many SCUBA
sources have had 24\,\micron\ counterparts identified in past surveys
\citep[e.g.][]{pope06,ivison07}. However, here we are talking about the
sources dominating the background at 850\,\um, which are fainter than the
typical sources detected by SCUBA. This is indicated by our simple luminosity function model, which suggests that a substantial number of low-luminosity sources are missing from the catalog at high redshifts. It also may be possible that some of
the high-redshift sources in our catalog have simply been
mis-identified as lower-redshift objects, but due to their relatively
small numbers, have little effect on the lower-redshift bins.

Either way, there is evidence that this study misses a portion of the
SFR history at the highest redshifts ($z > 1$), during which it is believed the
most massive galaxies were forming through powerful mergers. This is
consistent with the result of \citet{marsden09} who find the
contribution of \fidel\ sources to the CIB at 850\,\um\ to be
significantly low compared to the FIRAS measurements. They also find
that their estimate of the CIB generated at 850\,\um\ by sources with
$z > 1.2$ is low compared to the model of \citet{valiante09} which
otherwise fits the CIB at the \blast\ wavelengths.

\section{Conclusions}
We have studied the contribution of MIR galaxies selected at
24\,\um\ to the CIB and found that they resolve most, if not all, of
the radiation detected by the all-sky FIRAS and DIRBE surveys at the
peak of the far-infrared background.  We assigned redshifts to more
than 70\% of this population from existing redshift catalogs,
and developed a technique which uses the NIR
IRAC colors to statistically assign redshifts to the remainder of the catalog.
Using a stacking analysis to overcome the confusion noise arising from
the finite instrumental resolution, we study the composition of the
CIB as a function of redshift.  We find that 60\% of the CIB at
500\,\um\ is generated at redshifts $z>1.1$, while the
70\,\um\ background has a more recent origin, with 75\% of it being
generated at redshifts $z<1.1$.

By fitting SEDs to the mean flux densities at each redshift, we have
shown that, on average, this population is consistent with being
predominantly lower-luminosity LIRGs rather than the ULIRGs detected in SCUBA
surveys. It appears that a significant fraction of the submillimeter galaxies
known to exist at redshifts $z>2$ are missing from our analysis, and we
conjecture that they are simply below the 24\,\micron\ flux density
limit in our source catalog, or perhaps in some cases have had their redshifts
mis-identified. However, while such objects are important at high
redshifts, their contribution to the peak of the CIB is small compared
with the bulk of the MIR-selected galaxies at lower redshifts.

The evolution in the total comoving FIR luminosity density can be
evaluated using MIPS and \blast\ data directly, as these wavelengths
span the rest-frame FIR peak emission. This evolution can then be directly
converted into the star formation rate history. We find good agreement
with existing studies conducted at optical-UV wavelengths which have had
to rely on many more assumptions about physical source properties.

Our results confirm that dust obscuration dominates the history of
star formation of the Universe, with star formation rates that are
about three times larger than in the optical-UV, in the redshift range $0.1 <
z < 1$.

These results can be used to constrain models of luminosity densities
and galaxy formation as they directly probe the FIR energetics of the
Universe up to redshift $z \sim 3$.

\section{Acknowledgements}
We acknowledge the support of NASA through grant numbers NAG5-12785, NAG5-
13301, and NNGO-6GI11G, the NSF Office of Polar Programs, the Canadian Space Agency, the Natural Sciences and Engineering Research Council (NSERC) of Canada, and the UK Science and Technology Facilities Council (STFC). This work
is based in part on observations made with the
{\sl Spitzer Space Telescope}, which is operated by the
Jet Propulsion Laboratory, California Institute of
Technology under a contract with NASA. 

The authors are grateful to Herv\'e Dole for fruitful suggestions on the analysis,
and to Benjamin Magnelli for help with the FIDEL 24\,\um\ data.

\bibliography{ms}

\clearpage

\end{document}